\documentstyle[aps,prb,multicol]{revtex}

\title{Spin Properties of Low Density One-Dimensional Wires}
\author{K.~J. Thomas, J.~T. Nicholls, M. Pepper, W.~R. Tribe,
M.~Y. Simmons$^*$, and D.~A. Ritchie}
\address{Cavendish Laboratory,
Madingley Road,
Cambridge, CB3 OHE, United Kingdom}
\date{\today}
\begin{document}
\maketitle

\begin{abstract}
We report conductance measurements of a ballistic one-dimensional
(1D) wire defined in the lower two-dimensional electron gas
of a GaAs/AlGaAs double quantum well.
At low temperatures there is an additional structure
at $0.7(2e^2/h)$ in the conductance, which tends to  
$e^2/h$ as the electron density is decreased.
We find evidence for complete spin polarization in a weakly disorderd 
1D wire at zero magnetic field
through the observation of a conductance plateau at $e^2/h$, which
strengthens in an in-plane magnetic field and disappears with increasing
electron density. 
In all cases studied, with increasing temperature
structure occurs at $0.6(2e^2/h)$.
We suggest that the 0.7 structure is a many-body spin state 
excited out of, either
the spin-polarized electron gas at low densities, or the
spin-degenerate electron gas at high densities.

\end{abstract}
\pacs{73.61.-r, 73.23.Ad, 73.20.Dx, 73.23.-b}

\begin{multicols}{2}

One-dimensional (1D) semiconductor systems can be fabricated by a
variety of techniques. Some of the best quality devices, as determined by the
clarity of the quantized plateaus in the conductance characteristics,
are obtained by electrostatically squeezing a two-dimensional electron gas
(2DEG) at a GaAs/AlGaAs interface using a split-gate defined by electron-beam
lithography.\cite{wharam88}  The conductance, measured as a function of the
split-gate voltage, exhibits plateaus quantized at integer multiples
of $2e^2/h$, a result that is well understood as the adiabatic transmission
of spin-degenerate 1D subbands. However, after the last 1D subband has been
depopulated, an additional structure in the conductance has been measured at
$0.7(2e^2/h)$. One of the most revealing properties of this so-called 0.7 structure
is its evolution into the spin-split plateau at $e^2/h$ in a strong in-plane magnetic
field. There is also an enhancement of the $g$-factor as the 1D carrier density is
reduced. Both results suggest that there is a possible spin polarization
of the 1D electron gas at zero magnetic field.\cite{thomas96a}

Hartree-Fock calculations\cite{gold96} of electrons confined in a cylindrical
wire show that correlation effects are weak, and that at low electron densities
exchange interactions will drive a spontaneous spin polarization.
A spin polarization at zero magnetic field would give an extra
plateau in the conductance at $e^2/h$ rather than $0.7(2e^2/h)$.
To explain this discrepancy various
theories\cite{wang98,schmeltzer98,rejec99,flambaum00,spivak99} invoking
spin have been put forward.
Recent quantum Monte Carlo calculations\cite{malatesta99} show that in 1D the
paramagnetic state is always lower in energy than the ferromagnetic state,
so it is not clear whether the Hartree-Fock calculations are in
conflict with the Lieb-Mattis prediction\cite{lieb62}
that there is no ferromagnetic order in a 1D system.
The role of disorder in 1D systems is little understood,
but it has been shown\cite{andreev98} within mean-field theory
that for dimensions $d \leq 2$ a disordered
system may exhibit a partial spin polarization,
even though the system without disorder is paramagnetic.

The 0.7 structure is distinctly different from the conductance plateaus
measured\cite{yacoby96a} at multiples of $\alpha (2e^2/h)$
in long wires fabricated by overgrowth on a cleaved edge.
The renormalization of {\em all} the conductance plateaus
by the same factor, $\alpha \approx 0.85$,
has been interpreted\cite{alekseev98} as a reduction in the transmission probability,
due to poor impedance matching between the 
two-dimensional (2D) contacts and the 1D wire.
Wires created at the apex of a V-groove GaAs-AlGaAs heterojunction
also show\cite{kauf99} renormalized conductance plateaus due to poor 1D-2D coupling.
The structure at $0.7(2e^2/h)$ occurs in addition to the usual 
plateaus,\cite{thomas96a,thomas98b}
which are correctly quantized,
confirming that the source and drain reservoirs
are adiabatically connected to the 1D constriction.
The 0.7 structure has been observed in other
GaAs-based 1D wires, such as those created by wet
etching,\cite{krist98b} or by gating an undoped heterostructure.\cite{kane98,reilly00}
Zero-field spin splitting of 1D subbands has been observed\cite{grab99}
in constrictions fabricated from PbTe, a material that has a high dielectric
constant which is expected to suppress the electron-electron interactions.

In this paper we present new measurements of ballistic 1D
wires that remain relatively
free of impurity effects at electron densities
as low as $n \approx 3 \times 10^{10}$~cm$^{-2}$,
equal to a 1D electron density estimated to be
$n_{\rm{1D}} \approx 1.2 \times 10^{7}$~m$^{-1}$.
This has been achieved using coupled 1D wires,
where a negative voltage $V_{sg}$ on the split-gate creates
two parallel wires ($2 \times1$D)
out of a double quantum well system.
A voltage, $V_{mid}$, applied to a narrow midline
gate positioned in the center of the split gate,
allows charge to be shifted from one wire to the other in a
controllable fashion.\cite{thomas99a}
When only one of the two wires is conducting,
the conductance characteristics show cleaner plateaus
than 1D constrictions fabricated at a single heterojunction.\cite{thomas99a}
There is evidence of a plateau at $e^2/h$ as the 2D electron density $n$ is
decreased, and we show how the previously measured 0.7 structure
is related to this fully spin polarized state.

Two $2 \times1$D samples, A and B, were fabricated from double
quantum well wafers grown by molecular beam epitaxy, comprising
two 150~\AA-wide GaAs quantum wells separated by a 20~\AA\ (sample
A) or 25~\AA\ (sample B) Al$_{0.33}$Ga$_{0.67}$As barrier.
The double quantum well is doped both above and below using 2000~\AA\
of Si-doped ($1.2 \times 10^{17}$~cm$^{-3}$)
Al$_{0.33}$Ga$_{0.67}$As, offset by 600~\AA\ and 700~\AA\
Al$_{0.33}$Ga$_{0.67}$As spacer layers, respectively. The electron
density in each layer is approximately $1.3 \times
10^{11}$~cm$^{-2}$, with an average mobility of $1.45 \times
10^{6}$~cm$^{2}$/Vs. The wafers were processed into Hall bars with
AuNiGe Ohmic contacts that connect to both 2DEGs. 
Split-gates were defined by electron-beam lithography 
with the pattern shown in the Fig.~\ref{f:1} inset.
The split-gates have a length of 0.4~$\mu$m
and a gap width of 1.2~$\mu$m, 
and the midline gate has a width 0.4~$\mu$m.

Two-terminal differential conductance $G=dI/dV$ measurements
of the wires were carried out
in a dilution refrigerator using standard techniques.
To align the spin-degenerate plateaus at $2e^2/h$ and $4e^2/h$
series resistances of $R_s=700~\Omega$ 
and $R_s=750~\Omega$ have been subtracted from the zero-field
conductance traces of sample A and B; these $R_s$ values are
greater than the sample resistance when $V_{sg}=0$~V (345~$\Omega$
and 380~$\Omega$ for samples A and B).
When an in-plane magnetic field $B_{\parallel}$ is
applied parallel to the length of the split gate,
$R_s$ can become as high as 1.5~k$\Omega$;
however, at a given $B_{\parallel}$
the same resistance correction can be applied to all the
traces measured at different $V_{mid}$.
The 2D electron densities $n$,
which we use to characterize the 1D constrictions,
are measured from the number of edge states
that are transmitted by the wire
in the quantum Hall regime.

A previous investigation\cite{thomas99a} of sample B,
which is strongly coupled,
showed that matching the widths and electron densities of the two wires
brings them into resonance, forming symmetric and antisymmetric 1D subbands
that are separated by a gap that is larger than
the 2D value ($\Delta_{SAS} = 1.4$~meV).
In the measurements presented here, similar strongly coupled
wires are operated away from resonance,
in the regime where the top wire is pinched off.
Conduction proceeds only through the lower wire,
and the energy gap is unimportant.

Figure~\ref{f:1} shows the conductance characteristics
$G(V_{sg})$ of sample A at 50~mK,
obtained at different electron densities
as controlled by $V_{mid}$.
From left to right the $G(V_{sg})$ traces
are measured as $V_{mid}$ is varied from
-1.2~V to -5.4~V in steps of 0.3~V.
In this range of $V_{mid}$, the upper 1D wire is
completely depopulated and the conductance, quantized in units of
$2e^2/h$, originates from transport through the lower wire.
A clear 0.7 structure is present in all traces,
with a gradual shift of the structure to lower conductance
as the electron density is lowered with a more negative $V_{mid}$.
In previous measurements\cite{thomas98b}
of the 0.7 structure,
the electron density was varied using a back-gate,
and the conductance of the 0.7
structure decreased\cite{thomas98b} by 10\%
as the density was changed from 1.4 to $1.1 \times 10^{11}$~cm$^{-2}$.
The lowering of the conductance structure from $0.7(2e^2/h)$ to $0.53(2e^2/h)$
shown in Fig.~\ref{f:1} occurs when the density is reduced from
$1.3 \times 10^{11}$~cm$^{-2}$ to $3 \times 10^{10}$~cm$^{-2}$.
The midline gate has a stronger
effect on the electron density in the 1D channel than a backgate,
with the additional advantage that the 2D regions that constitute
the source and drain are not affected by $V_{mid}$.
More significantly, there is little degradation of the
quantization or flatness of the conductance plateaus
as the electron density $n$ is decreased to $3 \times 10^{10}$~cm$^{-2}$.
It is thought that the second parallel electron gas,
situated only 200~\AA\ away,
screens the electrons passing through
the entrance and exit of the constriction from impurities.

Figure~\ref{f:2} shows the temperature dependence of the
0.7 structure in sample A on a second cooldown.
For $V_{mid}=0$~V, where the density is $n=1.3 \times 10^{11}$cm$^{-2}$
the 0.7 structure drops down to $0.6(2e^2/h)$ as the temperature is
increased from 0.12~K to 1.9~K.
In contrast, when the density is
$n=3 \times 10^{10}$cm$^{-2}$ at $V_{mid}=-2.4$~V,
the structure at $0.55(2e^2/h)$
rises up to  $0.6(2e^2/h)$ as the temperature increases.
Therefore, the conductance tends to $0.6(2e^2/h)$
for temperatures greater than 2~K, whatever the electron density.
In a strong parallel magnetic field we have observed\cite{thomas96a,thomas98b}
the evolution of the 0.7 structure to a spin-split plateau at $e^2/h$.
If this high field state is warmed to 2-3~K
it also moves to $0.6(2e^2/h)$,\cite{thesis}
similar to the behavior at low densities seen
in the right hand side traces of Fig.~\ref{f:2}.
In both cases, the higher index plateaus do not rise with temperature,
showing that there is no change in the series resistance.

On taking $2 \times1$D devices to low electron densities,
some samples show cleaner conductance characteristics than others.
Figure~\ref{f:3}(a) shows the $G(V_{sg})$
characteristics for sample B at 80~mK,
where due to impurities
the conductance plateaus are not as flat as in sample A.
What is most surprising about this sample is that
when the electron density is reduced
below $4\times 10^{10}$cm$^{-2}$ ($V_{mid} < -0.94$~V)
a plateau at $e^2/h$ is observed at zero magnetic field;
this has been measured on three different cooldowns.
This zero-field $e^2/h$ plateau does not originate from
a 0.7 structure with decreasing density, as in sample A,
but appears suddenly as the electron density is decreased.
When the conductance measurements are repeated at 1.3~K,
see Fig.~\ref{f:3}(b), the $2e^2/h$ plateau becomes cleaner and the $e^2/h$
plateau develops into a strong structure at $0.6(2e^2/h)$.
This high temperature structure is present in all
traces in Fig.~\ref{f:3}(b),
even when there is no corresponding $e^2/h$ plateau at 80~mK.

The low density $e^2/h$ plateau in sample B
has been investigated for in-plane
magnetic fields $B_{\parallel}$ up to 16~T.
Figure~\ref{f:4} shows $G(V_{sg})$ traces at $V_{mid}=-1.06$~V
as $B_{\parallel}$ is increased in steps of 2~T.
The zero-field $e^2/h$ plateau strengthens and remains at $e^2/h$
as $B_{\parallel}$ is increased, indicating a spin splitting at
$B_{\parallel}=0$. Sample A, which is believed to have less
impurities than sample B at the same carrier density, does not
show a zero-field plateau at $e^2/h$. This suggests that the
spontaneous spin polarization in sample B is induced by weak
disorder, similar to the case studied in Ref.\onlinecite{andreev98}.

Our previous work\cite{thomas96a} showed that the 0.7 structure is due to
a possible spin polarization, which was accompanied by an enhancement of
the $g$-factor, both suggesting the importance of many-body interactions.
Here we have shown that as the electron concentration is reduced at low
temperatures, the 0.7 structure shifts down towards $0.5(2e^2/h)$,
suggesting that the system is moving towards a spin polarized ground state.
Whether or not such a spin polarized ground state is possible in
1D is still an open question. 
Lieb and Mattis have proved\cite{lieb62} that
in 1D the unpolarized state is always lower in energy than the polarized
state; real devices, however, are not strictly 1D because they have finite length
and non-zero width. 
Reimann {\it et al.}\cite{reimann99} have predicted that in finite 1D wires
there may be a spin-density wave (SDW), which could be a precursor to 
complete spin polarization.
The SDW may give rise to localized states at the
entrance and exit of the 1D wire, 
causing additional scattering that will reduce the conductance below $2e^2/h$.
We also report the case of a complete spin splitting at zero
magnetic field, which may come about
through the presence of disorder, 
though further clarification of this is required.

As shown by the Copenhagen group,\cite{krist98b}
it appears that the 0.7 structure is an excited state,
which at the lowest temperatures moves into the completely spin degenerate
state at $2e^2/h$.
We show here that at low temperatures the 0.7 structure
moves into the spin polarised $e^2/h$ state 
at low electron concentrations,
with weak disorder, or in an external in-plane magnetic field.
In all cases, the spin-split plateau at $0.5(2e^2/h)$
moves to a slightly higher conductance, typically $0.6(2e^2/h)$,
when the temperature is raised.
At present it is not clear why 
there is a 0.7 structure rather than plateau at $e^2/h$
at low temperatures and high densities,
there maybe a partial spin polarisation due
to a hybridization of the spin-up and spin-down state at 
higher temperatures.

In conclusion, we have shown that using just
one of the wires in a $2 \times1$D device,
the electron density in the constriction
can be taken to $3 \times 10^{10}$cm$^{-2}$,
without the conductance characteristics 
suffering so readily from impurity effects.
At low electron densities the 0.7 structure
moves towards $0.5(2e^2/h)$ at low temperatures.
We have also presented evidence for a spontaneous spin polarization,
possibly brought about by weak disorder, 
giving rise to a plateau at $e^2/h$. 
In all cases studied, 
for $T > 2$~K, structure is observed close to $0.6(2e^2/h)$. 
The temperature dependence suggests that the 0.7 structure 
is a many-body state that is excited
out of the spin polarized 1D electron gas at low densities,
or out of the spin-degenerate electron gas at high densities.

We thank the Engineering and Physical Sciences Research Council (UK) for
supporting this work, and JTN acknowledges an Advanced EPSRC Fellowship.

$^*$ Present address: Semiconductor Nanofabrication Facility,
School of Physics, University of New South Wales, Sydney 2052, Australia.


\begin{thebibliography}{10}

\bibitem{wharam88}
D.~A. Wharam, T.~J. Thornton, R. Newbury, M. Pepper, H. Ahmed, J.~E.~F. Frost,
D.~G. Hasko, D.~C. Peacock, D.~A. Ritchie, and G.~A.~C. Jones, J.\ Phys.\ C
{\bf 21},  L209  (1988).

\bibitem{thomas96a}
K.~J. Thomas, J.~T. Nicholls, M.~Y. Simmons, M. Pepper, D.~R. Mace, and D.~A.
Ritchie, Phys.\ Rev.\ Lett. {\bf 77},  135  (1996).

\bibitem{gold96}
A. Gold and L. Calmels, Phil.\ Mag.\ Lett. {\bf 74},  33  (1996).

\bibitem{wang98}
C.~K. Wang and K.-F. Berggren, Phys.\ Rev.\ B {\bf 57},  4552  (1998).

\bibitem{schmeltzer98}
D. Schmeltzer, E. Kogan, R. Berkovits, and M. Kaveh, Phil. Mag. B {\bf 77},
1189  (1998).

\bibitem{rejec99}
T. Rejec, A. Ram$\breve{\rm{s}}$ak, and J.~H. Jefferson, preprint,
cond-mat/9910399.

\bibitem{flambaum00}
V.~V. Flambaum and M.~Y. Kuchiev, Phys.\ Rev.\ B {\bf 61}, (2000),
cond-mat/9910415.

\bibitem{spivak99}
B. Spivak and F. Zhou, preprint, cond-mat/9911175.

\bibitem{malatesta99}
A. Malatesta and G. Senatore, preprint, cond-mat/9912342.

\bibitem{lieb62}
E.~H. Lieb and D. Mattis, Phys.\ Rev. {\bf 125},  164  (1962).

\bibitem{andreev98}
A.~V. Andreev and A. Kamenev, Phys.\ Rev.\ Lett. {\bf 81},  3199  (1998).

\bibitem{yacoby96a}
A. Yacoby, H.~L. Stormer, N.~S. Wingreen, L.~N. Pfeiffer, K.~W. Baldwin, and
K.~W. West, Phys.\ Rev.\ Lett. {\bf 77},  4612  (1996).

\bibitem{alekseev98}
A.~Y. Alekseev and V.~V. Cheianov, Phys.\ Rev.\ B {\bf 57},  6834  (1998).

\bibitem{kauf99}
D. Kaufman, Y. Berk, B. Dwir, A. Ridra, A. Palevski, and E. Kapon, Phys.\ Rev.\
B {\bf 59},  R10433  (1999).

\bibitem{thomas98b}
K.~J. Thomas, J.~T. Nicholls, N.~J. Appleyard, M. Pepper, M.~Y. Simmons, D.~R.
Mace, and D.~A. Ritchie, Phys.\ Rev.\ B {\bf 58},  4846  (1998).

\bibitem{krist98b}
A. Kristensen, M. Zaffalon, J. Hollingbery, C.~B. S{\o}renson, S.~M. Reimann,
P.~E. Lindelof, M. Michel, and A. Forchel, J.\ Appl.\ Phys. {\bf 83},  607
(1998).

\bibitem{kane98}
B.~E. Kane, G.~R. Facer, A.~S. Dzurak, N.~E. Lumpkin, R.~G. Clark, L.~N.
Pfeiffer, and K.~W. West, Appl.\ Phys.\ Lett. {\bf 72},  3506  (1998).

\bibitem{reilly00}
D.~J. Reilly, G.~R. Facer, A.~S. Dzurak, B.~E. Kane, R.~G. Clark, P.~J. Stiles,
J.~L. {O'Brien}, N.~E. Lumpkin, L.~N. Pfeiffer, and K.~W. West, preprint,
cond-mat/0001174.

\bibitem{grab99}
G. Grabecki, J. Wr$\acute{\rm{o}}$bel, T. Dietl, K. Byczuk, E. Papis, E.
Kami$\acute{\rm{n}}$ska, A. Piotrowska, G. Springholz, M. Pinczolits, and G.
Bauer, Phys.\ Rev.\ B {\bf 60},  R5133  (1999).

\bibitem{thomas99a}
K.~J. Thomas, J.~T. Nicholls, M.~Y. Simmons, W.~R. Tribe, A.~G. Davies, and M.
Pepper, Phys.\ Rev.\ B {\bf 59},  12252  (1999).

\bibitem{thesis}
K.~J. Thomas, Ph.D. thesis, University of Cambridge, 1997.

\bibitem{reimann99}
S.~M. Reimann, M. Koskinen, and M. Manninen, Phys.\ Rev.\ B {\bf 59},  1613
(1999).

\end{thebibliography}

\end{multicols}

\begin{figure}
\caption{{\em Inset:} Schematic plan and side views
of the submicron gates used to define a
$2\times1$D device in a double quantum well.
{\em Main:} The conductance characteristics $G(V_{sg})$
of sample A at $T=50$~mK. From left to right,
$V_{mid}$ is changed from
-1.2~V to -5.4~V in steps of 0.3~V. Dashed lines are drawn
at $0.7(2e^2/h)$ and $0.5(2e^2/h)$.}
\label{f:1}
\end{figure}

\begin{figure}
\caption{Temperature dependence of $G(V_{sg})$ 
characteristics of sample A on a different cooldown.
The different sets of traces correspond
to $V_{mid}=$~0, -1.2, -1.8, and -2.4~V (left to right). 
At $T=2$~K structure occurs at $0.6(2e^2/h)$ for all densities.} 
\label{f:2}
\end{figure}

\begin{figure}
\caption{Zero-field conductance characteristics
$G(V_{sg})$ of sample B at (a) $T=80$~mK, and (b) $T=1.3$~K.
At both temperatures $V_{mid}$ is decreased (left to right)
from -0.86~V to -1.0~V, in steps of 0.01~V.
The plateau at $e^2/h$ measured at 80~mK
moves to 0.6$(2e^2/h)$ at higher temperatures.}
\label{f:3}
\end{figure}

\begin{figure}
\caption{Conductance characteristics $G(V_{sg})$ of sample B
at 80~mK and $V_{mid}=$-1.06~V.
From left to right the in-plane magnetic field $B_{\parallel}$
is increased from 0 to 16~T in steps of 2~T;
the plateau at $e^2/h$ strengthens with $B_{\parallel}$.
For clarity, successive traces have been horizontally offset by 114~mV.
The zero-field pinch-off characteristics are slightly different from those
in Fig.~\ref{f:3}(a), as the two measurements were taken four days
apart.}
\label{f:4}
\end{figure}

\end{document}